\documentclass{qjmam}

\usepackage{amsfonts,amsmath,amssymb,bm}

\startpage{1}
\yr{0}
\vol{0}
\issue{0}

\newtheorem{definition}{Definition}[section]
\newtheorem{lemma}[definition]{Lemma}
\newtheorem{Theorem}[definition]{Theorem}

\DeclareMathAlphabet\mathbit
    \encodingdefault\rmdefault\bfdefault\itdefault
\DeclareOldFontCommand{\bi}{\normalfont\bfseries\itshape}{\mathbit}

\newcommand{\be}{\begin{equation}}
\newcommand{\ee}{\end{equation}}

\def\fakebold#1{\relax\ifvmode\leavevmode\fi%
\ifmmode%
\setbox0=\hbox{$#1$}%
\else%
\setbox0=\hbox{#1}%
\fi%
\kern-.02em\copy0 \kern-\wd0%
\kern .04em\copy0 \kern-\wd0%
\kern-.0125em\raise.02em\box0%
}%

\begin{document}

\title[{Rotational~elasticity}] {ROTATIONAL ELASTICTY}

\author[C.~G.~B\"ohmer \etal] {C.~G.~B\"ohmer,
R.~J.~Downes \and D.~Vassiliev}

\address{Department of Mathematics and Institute of Origins,
University College London,\\
Gower Street, London, WC1E 6BT, United Kingdom}

\received{\recd date. \revd date}

\maketitle

\eqnobysec

\begin{abstract}
We consider an infinite 3-dimensional elastic continuum whose
material points experience no displacements, only rotations. This
framework is a special case of the Cosserat theory of elasticity.
Rotations of material points are described mathematically by
attaching to each geometric point an orthonormal basis which gives a
field of orthonormal bases called the coframe. As the dynamical
variables (unknowns) of our theory we choose the coframe and a
density. We write down the general dynamic variational functional
for our rotational theory of elasticity, assuming our material to be
physically linear but the kinematic model geometrically nonlinear.
Allowing geometric nonlinearity is natural when dealing with
rotations because rotations in dimension 3 are inherently nonlinear
(rotations about different axes do not commute) and because there is
no reason to exclude from our study large rotations such as full
turns. The main result of the paper is an explicit construction of a
class of time-dependent solutions which we call plane wave
solutions; these are travelling waves of rotations. The existence of
such explicit closed form solutions is a nontrivial fact given that
our system of Euler--Lagrange equations is highly nonlinear. In the
last section we consider a special case of our rotational theory of
elasticity which in the stationary setting (harmonic time dependence
and arbitrary dependence on spatial coordinates) turns out to be
equivalent to a pair of massless Dirac equations.
\end{abstract}


\section{Introduction}
\label{Introduction}

We work in 3-dimensional Euclidean space and
view it as an elastic continuum whose material
points can experience no displacements, only rotations, with
rotations of different material points being totally independent.
Rotations of material points of the 3-dimensional elastic continuum
are described mathematically by attaching to each geometric point an
orthonormal basis. This gives a field of orthonormal bases called
the \emph{coframe}.

The purpose of our paper is to develop a theory of elasticity on
rotations, i.e.~a theory of elasticity in which the coframe plays
the role of the dynamical variable (unknown quantity). Recall that
in classical elasticity the vector field of displacements is the
dynamical variable.

Our motivation for studying such a seemingly exotic problem comes
from three main sources.

The first source is \emph{Cosserat elasticity}. In 1909 the Cosserat
brothers proposed a theory of elasticity \cite{Co} which generalised
classical elasticity by giving each material point rotational
degrees of freedom. Cosserat elasticity has since become an accepted
part of solid mechanics, though for most real life materials effects
resulting from rotations of material points are small compared to
effects resulting from displacements. From a purely mathematical
point of view classical elasticity and rotational elasticity are two
limit cases of Cosserat elasticity. One of these limit cases,
classical elasticity, has been extensively studied so it seems
natural to examine now the other limit case.

The second source is \emph{teleparallelism} (=~absolute parallelism
=~fernparallelismus), a subject promoted by A.~Einstein and
\'E.~Cartan \cite{MR543192,MR2276051,unzicker-2005-} in the late 1920s.
The idea of rotating material points lies at the heart of
teleparallelism and can easily be traced back to Cosserat elasticity:
when in 1922
Cartan started developing what eventually became modern differential geometry
he acknowledged~\cite{cartan1}
that he drew inspiration from the `beautiful' work
of the Cosserat brothers.
The relationship between Cosserat elasticity
and teleparallelism is examined in detail in the review paper
\cite{cartantorsionreview}.

The third source is the theory of liquid crystals and, in
particular, the concept of an \emph{Ericksen fluid}. According to
\cite{ericksen_twist_waves}, in a liquid crystal one can observe
`orientation waves which propagate, inducing little or no motion of
the fluid' and Ericksen's mathematical model is the natural way of
describing this phenomenon. The only difference between Ericksen's
model and ours is that in Ericksen's model one attaches to each
geometric point a single unit vector rather than an orthonormal
basis, as we do.

Our paper has the following structure.
In Section~\ref{Setting the playing field}
we define our dynamical variables (unknowns of our theory),
in Section~\ref{Kinetic energy}
we write down the kinetic energy
and Section~\ref{Potential energy}
we write down the potential energy.
The Lagrangian of rotational elasticity is written down in
Section~\ref{Lagrangian of rotational elasticity}.
In Section~\ref{Reformulating the problem in the language of spinors}
we reformulate our model in the language of spinors
and in Section~\ref{Euler--Lagrange equation}
we discuss the corresponding Euler--Lagrange equation.
In Section~\ref{Plane wave solutions}
we construct an explicit class of solutions which we call plane
wave solutions; this construction is summarised in
Theorem~\ref{Theorem 1} which is the
main result of our paper.
Finally, in Section~\ref{The massless Dirac equation}
we compare our model with the massless Dirac equation.

\section{Setting the playing field}
\label{Setting the playing field}

We work in Euclidean space $\mathbb{R}^3$ equipped with Cartesian
coordinates $x^\alpha$, $\alpha=1,2,3$, and standard Euclidean metric.
We denote time by $x^0$. Partial differentiation in $x^0$ and $x^\alpha$,
$\alpha=1,2,3$, is denoted by $\partial_0$ and $\partial_\alpha$
respectively.

The coframe $\vartheta$ is a triple of orthonormal covector fields
$\vartheta^j$, $j=1,2,3$, in $\mathbb{R}^3$. Each covector field
$\vartheta^j$ can be written more explicitly as
$\vartheta^j{}_\alpha$ where the tensor index $\alpha=1,2,3$
enumerates the components. The orthonormality condition for the
coframe can be represented as a single tensor identity
\begin{equation}
\label{constraint for coframe}
g=\delta_{jk}\vartheta^j\otimes\vartheta^k
\end{equation}
where $\delta$ is the Kronecker delta and
$g=g_{\alpha\beta}=\delta_{\alpha\beta}$
is the Euclidean metric.
For the sake of clarity we repeat formula~(\ref{constraint for coframe})
giving tensor indices explicitly and performing summation over frame
indices explicitly:
$
\delta_{\alpha\beta}
=\vartheta^1{}_\alpha\vartheta^1{}_\beta
+\vartheta^2{}_\alpha\vartheta^2{}_\beta
+\vartheta^3{}_\alpha\vartheta^3{}_\beta
$
where $\alpha$ and $\beta$ run through the values $1,2,3$.
We view the identity (\ref{constraint for coframe}) as a kinematic
constraint: the covector fields
$\vartheta^j$ are chosen so that they
satisfy~(\ref{constraint for coframe}), which leaves us with three real
degrees of freedom at every point of $\mathbb{R}^3$.

We work only with coframes which have positive orientation,
i.e.~which satisfy the condition
\begin{equation}
\label{orientation of coframe}
\det\vartheta^j{}_\alpha=+1>0.
\end{equation}

If one views $\vartheta^j{}_\alpha$ as a $3\times3$ real
matrix-function, then conditions (\ref{constraint for coframe}) and
(\ref{orientation of coframe}) mean that this matrix-function is
special orthogonal. Thus, the coframe can be thought of as a field
of special orthogonal matrices.

As dynamical variables in our model we choose the coframe
$\vartheta$ and a positive density $\rho$. Our coframe and density
are functions of Cartesian coordinates $x^\alpha$, $\alpha=1,2,3$,
as well as of time $x^0$.
At a physical level, making the density $\rho$ a
dynamical variable means that we view our continuum more like a
fluid rather than a solid. In other words, we allow the material to
redistribute itself so that it finds its equilibrium density distribution.
Observe that the total number of real dynamical degrees of freedom
contained in the coframe $\vartheta$ and positive density $\rho$
is four, exactly as in a two-component complex-valued spinor field.

Note that there is nothing wrong in taking a prescribed density (as
opposed to a density which is a dynamical variable): the theory one
gets is very similar to the one described in the current paper and
most formulae carry through with minimal changes.

Below is the list of the main assumptions on which our model will be
based.

\emph{Assumption 1: our model is geometrically nonlinear.} This
means that we do not linearise rotations and we do not linearise the
density. In other words, we allow our material points to experience
full turns and we allow our density to experience changes comparable
to the density itself.

\emph{Assumption 2: our material is physically linear.} This means
that our potential energy is chosen to be quadratic in torsion (the
latter serves as the measure of rotational deformations, see
subsection \ref{Measuring rotational deformations}). Note that
physical linearity does not contradict geometric nonlinearity:
locally (in space and time) material points ``do not know'' that
they may eventually experience full rotations and the density ``does
not know'' that it may eventually experience a change comparable to
its current value.

\emph{Assumption 3: our material is homogeneous and isotropic.}
Homogeneity means that physical properties of the material are the
same at all points of our continuum and isotropy means that there
are no preferred directions.

\emph{Assumption 4: our model is invariant under rigid rotations of the coframe.}
By a rigid rotation of the coframe we understand the transformation
\begin{equation}
\label{rigid rotations of the coframe}
\vartheta^j\mapsto O^j{}_k\vartheta^k
\end{equation}
where $O^j{}_k$ is a \emph{constant} special orthogonal matrix. The
thinking here is that when we attach an orthonormal basis to each
geometric point of our continuum there is no reason to associate one
particular direction with $\vartheta^1$, another with $\vartheta^2$
and a third with $\vartheta^3$. What matters is how these directions
change when we move from one point to another, i.e.~how orthonormal
bases at different points differ relative to each other. A rigid
rotation of the coframe means that we simultaneously rotate all our
orthonormal bases by the same angle around the same axis. We view
rigid rotations of the coframe as gauge transformations and assume
that our model does not feel them. See also \cite{lazar}
for a detailed exposition of gauge theory
for problems similar to the ones considered in our paper.

\section{Kinetic energy}
\label{Kinetic energy}

Kinetic energy is given by the formula
\begin{equation}
\label{Kinetic energy equation 1}
K(x^0)=c^\mathrm{kin}\int\|\omega\|^2\rho\,dx^1dx^2dx^3
\end{equation}
where $c^\mathrm{kin}$ is some positive constant and
$\omega$ is the (pseudo)vector of angular velocity
\begin{equation}
\label{definition of angular velocity}
\omega=\frac12*(\delta_{jk}\vartheta^j\wedge\partial_0\vartheta^k).
\end{equation}
Here $\wedge$ is the exterior product and $\,*\,$ is the Hodge star
(\ref{definition of the Hodge star}).

In writing the formula for kinetic energy
(\ref{Kinetic energy equation 1}) we think
of each material point as a uniform ball possessing a moment of
inertia and without a preferred axis of rotation.

We give for reference a more explicit version of the formula
for angular velocity
(\ref{definition of angular velocity}):
\begin{equation}
\label{Kinetic energy equation 3}
\omega_\alpha=\frac12
\sum_{j=1}^3
\begin{pmatrix}
\vartheta^j{}_2\partial_0\vartheta^j{}_3-\vartheta^j{}_3\partial_0\vartheta^j{}_2\\
\vartheta^j{}_3\partial_0\vartheta^j{}_1-\vartheta^j{}_1\partial_0\vartheta^j{}_3\\
\vartheta^j{}_1\partial_0\vartheta^j{}_2-\vartheta^j{}_2\partial_0\vartheta^j{}_1
\end{pmatrix}.
\end{equation}

\section{Potential energy}
\label{Potential energy}

\subsection{Measuring rotational deformations}
\label{Measuring rotational deformations}

In order to write down the formula for the potential energy we need to measure
deformations caused by rotations of the coframe. More specifically, we
need to measure deformations caused by the fact that at different
points the coframe is oriented differently. Obvious candidates for a
measure of deformations are the rank two tensors
\begin{equation}
\label{Measuring rotational deformations equation 1}
K^j:=\partial\vartheta^j,\qquad j=1,2,3,
\end{equation}
or, in more explicit form,
$K^j{}_{\alpha\beta}:=\partial_\alpha\vartheta^j{}_\beta$.
The problem is that taken separately the three rank two tensors
(\ref{Measuring rotational deformations equation 1})
are not invariant under rigid rotations of the coframe
(\ref{rigid rotations of the coframe}).
The natural
way of forming a truly invariant object is to make one rank three
tensor out of the three rank two tensors $K^j$ according to the
formula
\begin{equation}
\label{Measuring rotational deformations equation 3}
K:=\delta_{jk}\vartheta^j\otimes K^k
=\delta_{jk}\vartheta^j\otimes\partial\vartheta^k.
\end{equation}
The rank three tensor $K$ is invariant under rigid rotations of the coframe
(\ref{rigid rotations of the coframe})
and, moreover, the individual rank two tensors $K^j$ can be recovered from $K$ as
$K^j{}_{\gamma\delta}=\vartheta^{j\alpha}K_{\alpha\gamma\delta}$
so there is no loss of information.

Let us examine the symmetries of the tensor $K$. Observe that
formula (\ref{constraint for coframe}) implies
\[
0=\partial_\alpha g_{\beta\gamma}
=\partial_\alpha(\delta_{jk}\vartheta^j{}_\beta\vartheta^k{}_\gamma)
=\delta_{jk}(\partial_\alpha\vartheta^j{}_\beta)\vartheta^k{}_\gamma
+\delta_{jk}\vartheta^j{}_\beta(\partial_\alpha\vartheta^k{}_\gamma)
=K_{\gamma\alpha\beta}+K_{\beta\alpha\gamma}
\]
which means that the rank three tensor $K$ is antisymmetric in the
first and third indices,
\begin{equation}
\label{Measuring rotational deformations equation 4}
K_{\gamma\alpha\beta}=-K_{\beta\alpha\gamma}.
\end{equation}

Now, let us introduce another rank three tensor
\begin{equation}
\label{Measuring rotational deformations equation 5}
T:=\delta_{jk}\vartheta^j\otimes d\vartheta^k
\end{equation}
where $d$ stands for the exterior derivative. The tensor
(\ref{Measuring rotational deformations equation 5})
is obviously antisymmetric in the second and third indices
\begin{equation}
\label{Measuring rotational deformations equation 6}
T_{\alpha\beta\gamma}=-T_{\alpha\gamma\beta}
\end{equation}
and is expressed via our original deformation tensor
(\ref{Measuring rotational deformations equation 3})
as
\begin{equation}
\label{Measuring rotational deformations equation 7}
T_{\alpha\beta\gamma}=K_{\alpha\beta\gamma}-K_{\alpha\gamma\beta}.
\end{equation}
Formulae
(\ref{Measuring rotational deformations equation 7})
and
(\ref{Measuring rotational deformations equation 4})
imply
\begin{eqnarray*}
T_{\alpha\beta\gamma}&=&K_{\alpha\beta\gamma}+K_{\beta\gamma\alpha},\\
T_{\gamma\alpha\beta}&=&K_{\gamma\alpha\beta}+K_{\alpha\beta\gamma},\\
T_{\beta\gamma\alpha}&=&K_{\beta\gamma\alpha}+K_{\gamma\alpha\beta}
\end{eqnarray*}
where the last two identities were obtained from the first one by a cyclic
relabelling of tensor indices.
Adding up the first and second identities and subtracting the third
one we get
\begin{equation}
\label{Measuring rotational deformations equation 8}
K_{\alpha\beta\gamma}
=\frac12(T_{\alpha\beta\gamma}+T_{\gamma\alpha\beta}-T_{\beta\gamma\alpha})
=\frac12(T_{\alpha\beta\gamma}+T_{\beta\alpha\gamma}+T_{\gamma\alpha\beta})
\end{equation}
(here we also used (\ref{Measuring rotational deformations equation 6})).
Note that the argument carried out above is a rephrasing of the
standard argument that for a metric compatible affine connection
contortion can be expressed via torsion, see
subsection~7.2.6 in \cite{nakahara}.

Formulae
(\ref{Measuring rotational deformations equation 7})
and
(\ref{Measuring rotational deformations equation 8})
show that the tensors
$K$ and $T$ are expressed via each other so either of them can be used
as a measure of rotational deformations. We choose to use the tensor $T$ because
it has a clear geometric meaning: it is the torsion of the teleparallel connection
generated by the coframe $\vartheta$,
see Appendix A of \cite{MR2573111} for a concise exposition.
An additional advantage
of using the tensor $T$ is that the definition
(\ref{Measuring rotational deformations equation 5})
of this tensor does not require the use of covariant
derivatives so it works when the metric $g$ appearing in formula
(\ref{constraint for coframe})
is not assumed to
be Euclidean. The latter was important for Einstein and Cartan who arrived
at the mathematical model similar to the one described in
in our paper coming from general relativity.
Recall that in general relativity the metric plays the role of dynamical
variable so for someone with a relativistic background assuming the metric to be
Euclidean (i.e.~space to be flat) is unnatural.

Starting from Einstein's works \cite{unzicker-2005-} torsion is
traditionally used as a measure of deformations when modelling
elastic continua with rotations. We shall follow this tradition and
construct our potential energy as a function(al) of $T$. However,
before writing down the formula for potential energy we will
simplify matters by using the fact that we are working in 3D (our
previous arguments were dimension-independent).

Applying the Hodge star
(\ref{definition of the Hodge star})
in the second and third indices we switch
from the original torsion tensor $T$ to the tensor
\begin{equation}
\label{Measuring rotational deformations equation 9}
\overset{*}T_{\alpha\beta}:=
\frac12T_\alpha{}^{\gamma\delta}\varepsilon_{\gamma\delta\beta}.
\end{equation}
Of course, the tensor $T$ can be recovered from $\overset{*}T$ as
\begin{equation}
\label{Measuring rotational deformations equation 10}
T_{\alpha\beta\gamma}=
\overset{*}T_\alpha{}^\delta\varepsilon_{\delta\beta\gamma}.
\end{equation}

Formulae
(\ref{Measuring rotational deformations equation 9})
and
(\ref{Measuring rotational deformations equation 10})
show that the tensors $T$ and $\overset{*}T$ are expressed via each other so either
of them can be used as a measure of rotational deformations. We
choose to use the tensor $\overset{*}T$ because it has lower rank,
two instead of three.

Formulae
(\ref{Measuring rotational deformations equation 5})
and
(\ref{Measuring rotational deformations equation 9})
imply
\begin{equation}
\label{Measuring rotational deformations equation 11}
\overset{*}T=\delta_{jk}\vartheta^j\otimes *d\vartheta^k
=\delta_{jk}\vartheta^j\otimes\operatorname{curl}\vartheta^k.
\end{equation}
We see that $\overset{*}T$ is a rank two tensor without any
symmetries and with arbitrary trace. This is the tensor we will be
using as a measure of rotational deformations when writing down the
formula for potential energy. The tensor $\overset{*}T$ is sometimes
called the \emph{dislocation density tensor}  \cite{kleinert} .

We give for reference a more explicit version of formula
(\ref{Measuring rotational deformations equation 11}):
\begin{equation}
\label{Measuring rotational deformations equation 12}
\overset{*}T_{\alpha\beta}=
\sum_{j=1}^3
\begin{pmatrix}
\vartheta^j{}_1\partial_2\vartheta^j{}_3-\vartheta^j{}_1\partial_3\vartheta^j{}_2&
\vartheta^j{}_1\partial_3\vartheta^j{}_1-\vartheta^j{}_1\partial_1\vartheta^j{}_3&
\vartheta^j{}_1\partial_1\vartheta^j{}_2-\vartheta^j{}_1\partial_2\vartheta^j{}_1\\
\vartheta^j{}_2\partial_2\vartheta^j{}_3-\vartheta^j{}_2\partial_3\vartheta^j{}_2&
\vartheta^j{}_2\partial_3\vartheta^j{}_1-\vartheta^j{}_2\partial_1\vartheta^j{}_3&
\vartheta^j{}_2\partial_1\vartheta^j{}_2-\vartheta^j{}_2\partial_2\vartheta^j{}_1\\
\vartheta^j{}_3\partial_2\vartheta^j{}_3-\vartheta^j{}_3\partial_3\vartheta^j{}_2&
\vartheta^j{}_3\partial_3\vartheta^j{}_1-\vartheta^j{}_3\partial_1\vartheta^j{}_3&
\vartheta^j{}_3\partial_1\vartheta^j{}_2-\vartheta^j{}_3\partial_2\vartheta^j{}_1
\end{pmatrix}.
\end{equation}

\subsection{Irreducible decomposition of rotational deformations}
\label{Irreducible decomposition of rotational deformations}

Recall the logic of classical linear elasticity \cite{MR884707}:
after identifying the measure of deformation one decomposes it into
irreducible pieces. We follow this logic by decomposing the tensor
$\overset{*}T$ into irreducible pieces. The construction presented
below is similar to \cite{MR884707}, the only difference being that
instead of a symmetric rank two tensor, strain, we deal with a rank
two tensor, $\overset{*}T$, without any symmetries.

Decomposing the rank two tensor $\overset{*}T$ into irreducible pieces
means the following. We fix a point in $\mathbb{R}^3$ and at this
point look at all real rank two tensors $P$.
Such tensors can be viewed
as elements of a real 9-dimensional vector space $V$ equipped with
inner product
\begin{equation}
\label{Irreducible decomposition of rotational deformations equation 1}
(P,Q)_V:=P_{\alpha\beta}Q^{\alpha\beta}
\end{equation}
and corresponding norm
\begin{equation}
\label{Irreducible decomposition of rotational deformations equation 2}
\|P\|_V=\sqrt{(P,P)_V}=\sqrt{P_{\alpha\beta}P^{\alpha\beta}}\,.
\end{equation}
Let us now examine what happens when we rotate our Cartesian
coordinate system~$x^\alpha$, i.e.~when we perform a linear change
of coordinates preserving the metric $g_{\alpha\beta}$ and
orientation. The components of our tensors $P_{\alpha\beta}$
change in a particular way under rotations of the coordinate system,
so we get an action of the group $\mathrm{SO}(3)$ on the vector
space $V$. Looking for irreducible pieces of torsion means
identifying subspaces of $V$ which are invariant under the action of
the group $\mathrm{SO}(3)$, i.e.~which map into themselves,
and which do not contain smaller nontrivial invariant subspaces.

In our case the invariant subspaces are obvious. These are
\begin{itemize}
\item
the 1-dimensional subspace of real rank two tensors proportional to
the metric,
\item
the 3-dimensional subspace of real antisymmetric rank two tensors and
\item
the 5-dimensional subspace of real symmetric trace-free rank two tensors.
\end{itemize}
These three subspaces are clearly irreducible and mutually
orthogonal in the inner
product~(\ref{Irreducible decomposition of rotational deformations equation 1}).

Our rank two tensor $\overset{*}T$ can now be written as a sum of
three irreducible pieces
\begin{equation}
\label{Irreducible decomposition of rotational deformations equation 3}
\overset{*}T=
\overset{*}T{}^\mathrm{ax}+\overset{*}T{}^\mathrm{vec}+\overset{*}T{}^\mathrm{ten}
\end{equation}
where
\begin{eqnarray}
\label{Irreducible decomposition of rotational deformations equation 4}
\overset{*}T{}^\mathrm{ax}_{\alpha\beta}&:=&
\frac{\overset{*}T{}^\gamma{}_\gamma}3\,g_{\alpha\beta}\,,\\
\label{Irreducible decomposition of rotational deformations equation 5}
\overset{*}T{}^\mathrm{vec}_{\alpha\beta}&:=&
\frac{\overset{*}T{}_{\alpha\beta}-\overset{*}T{}_{\beta\alpha}}2\,,\\
\label{Irreducible decomposition of rotational deformations equation 6}
\overset{*}T{}^\mathrm{ten}_{\alpha\beta}&:=&
\overset{*}T_{\alpha\beta}
-\overset{*}T{}^\mathrm{ax}_{\alpha\beta}
-\overset{*}T{}^\mathrm{vec}_{\alpha\beta}=
\frac{\overset{*}T{}_{\alpha\beta}+\overset{*}T{}_{\beta\alpha}}2
-\frac{\overset{*}T{}^\gamma{}_\gamma}3\,g_{\alpha\beta}\,.
\end{eqnarray}
We label the irreducible pieces
(\ref{Irreducible decomposition of rotational deformations equation 4}),
(\ref{Irreducible decomposition of rotational deformations equation 5})
and
(\ref{Irreducible decomposition of rotational deformations equation 6})
by the adjectives
\emph{axial}, \emph{vector} and \emph{tensor} respectively,
which is terminology traditional in
alternative theories of gravity \cite{cartantorsionreview}.

\subsection{Formula for potential energy}
\label{Formula for potential energy}

Following the logic of classical linear elasticity \cite{MR884707} we
now write down the explicit formula for potential energy:
\begin{equation}
\label{Formula for potential energy equation 1}
P(x^0)=\int
\bigl(
c^\mathrm{ax}\|\overset{*}T{}^\mathrm{ax}\|_V^2+
c^\mathrm{vec}\|\overset{*}T{}^\mathrm{vec}\|_V^2+
c^\mathrm{ten}\|\overset{*}T{}^\mathrm{ten}\|_V^2
\bigr)\rho\,dx^1dx^2dx^3
\end{equation}
where $c^\mathrm{ax}$, $c^\mathrm{vec}$ and $c^\mathrm{ten}$ are some
nonnegative constants
(elastic moduli), not all zero,
and $\|\,\cdot\,\|_V$ is the norm
(\ref{Irreducible decomposition of rotational deformations equation 2}).
Comparing our formula
(\ref{Formula for potential energy equation 1})
with formula (4.3) from \cite{MR884707}
we see a difference between classical
and rotational elasticity:
classical elasticity involves two elastic moduli
whereas rotational elasticity involves three.
The extra elastic modulus $c^\mathrm{vec}$ is needed because the tensor
$\overset{*}T$ which we use as measure of rotational deformations is
not necessarily symmetric.

Formula (\ref{Formula for potential energy equation 1}) is the one
traditionally used in teleparallelism. This formula already appears
in the original papers of Einstein \cite{unzicker-2005-}, though for some
reason\footnote{The reason could be that Einstein was primarily
interested in providing a geometric interpretation of
electromagnetism and might have felt that the axial term would not
contribute to the electromagnetic field.} Einstein did not include
the axial term $c^\mathrm{ax}\|\overset{*}T{}^\mathrm{ax}\|^2$.
Subsequent authors always used three terms,
see, for example, formula (26) in \cite{cartantorsionreview}.


\subsection{Simplifying the formula for potential energy}
\label{Simplifying the formula for potential energy}

Let us introduce the (pseudo)scalar
\begin{equation}
\label{Simplifying the formula for potential energy equation 1}
f:=\overset{*}T{}^\alpha{}_\alpha
\end{equation}
and the vector
\begin{equation}
\label{Simplifying the formula for potential energy equation 2}
v_\alpha:=\overset{*}T{}^{\beta\gamma}\varepsilon_{\beta\gamma\alpha}\,.
\end{equation}
Formulae
(\ref{Irreducible decomposition of rotational deformations equation 2})--(\ref{Irreducible decomposition of rotational deformations equation 6}),
(\ref{Simplifying the formula for potential energy equation 1})
and
(\ref{Simplifying the formula for potential energy equation 2})
imply
\begin{equation}
\label{Simplifying the formula for potential energy equation 3}
\|\overset{*}T{}^\mathrm{ax}\|_V^2=\frac13f^2,
\end{equation}
\begin{equation}
\label{Simplifying the formula for potential energy equation 4}
\|\overset{*}T{}^\mathrm{vec}\|_V^2=\frac12\|v\|^2,
\end{equation}
\begin{equation}
\label{Simplifying the formula for potential energy equation 5}
\|\overset{*}T{}^\mathrm{ten}\|_V^2=
\|\overset{*}T{}\|_V^2-\frac13f^2-\frac12\|v\|^2.
\end{equation}
Substituting
(\ref{Measuring rotational deformations equation 12})
into
(\ref{Simplifying the formula for potential energy equation 1})
and
(\ref{Simplifying the formula for potential energy equation 2})
we get more explicit formulae for $f$ and $v$:
\begin{equation}
\label{Simplifying the formula for potential energy equation 6}
f=\sum_{j=1}^3
(\vartheta^j{}_1\partial_2\vartheta^j{}_3-\vartheta^j{}_1\partial_3\vartheta^j{}_2
+\vartheta^j{}_2\partial_3\vartheta^j{}_1-\vartheta^j{}_2\partial_1\vartheta^j{}_3
+\vartheta^j{}_3\partial_1\vartheta^j{}_2-\vartheta^j{}_3\partial_2\vartheta^j{}_1),
\end{equation}
\begin{equation}
\label{Simplifying the formula for potential energy equation 7}
v_\alpha=\sum_{j=1}^3
\begin{pmatrix}
\vartheta^j{}_2\partial_1\vartheta^j{}_2-\vartheta^j{}_2\partial_2\vartheta^j{}_1
-\vartheta^j{}_3\partial_3\vartheta^j{}_1+\vartheta^j{}_3\partial_1\vartheta^j{}_3
\\
\vartheta^j{}_3\partial_2\vartheta^j{}_3-\vartheta^j{}_3\partial_3\vartheta^j{}_2
-\vartheta^j{}_1\partial_1\vartheta^j{}_2+\vartheta^j{}_1\partial_2\vartheta^j{}_1
\\
\vartheta^j{}_1\partial_3\vartheta^j{}_1-\vartheta^j{}_1\partial_1\vartheta^j{}_3
-\vartheta^j{}_2\partial_2\vartheta^j{}_3+\vartheta^j{}_2\partial_3\vartheta^j{}_2
\end{pmatrix}.
\end{equation}

Substituting formulae
(\ref{Simplifying the formula for potential energy equation 1})--(\ref{Simplifying the formula for potential energy equation 3})
into formula
(\ref{Formula for potential energy equation 1})
we get
\begin{equation}
\label{Simplifying the formula for potential energy equation 8}
P(x^0)=\int
\left(
\frac{c^\mathrm{ax}-c^\mathrm{ten}}3f^2
+
\frac{c^\mathrm{vec}-c^\mathrm{ten}}2\|v\|^2
+
c^\mathrm{ten}\overset{*}T{}_{\alpha\beta}\overset{*}T{}^{\alpha\beta}
\right)\rho\,dx^1dx^2dx^3\,.
\end{equation}
The advantage of writing potential energy in the form
(\ref{Simplifying the formula for potential energy equation 8})
is that the geometric quantities $f$, $v$ and $\overset{*}T$
appearing in this formula have relatively compact explicit
representations
(\ref{Simplifying the formula for potential energy equation 6}),
(\ref{Simplifying the formula for potential energy equation 7})
and
(\ref{Measuring rotational deformations equation 12}).

\section{Lagrangian of rotational elasticity}
\label{Lagrangian of rotational elasticity}

We combine our potential energy
(\ref{Simplifying the formula for potential energy equation 8})
and kinetic energy
(\ref{Kinetic energy equation 1})
in forming the action (variational function) of dynamic
rotational elasticity
\begin{equation}
\label{Lagrangian of rotational elasticity equation 1}
S(\vartheta,\rho)=\int(P(x^0)-K(x^0))dx^0
=\int L(\vartheta,\rho)\,dx^0dx^1dx^2dx^3
\end{equation}
where
\begin{equation}
\label{Lagrangian of rotational elasticity equation 2}
L(\vartheta,\rho)=
\left(
\frac{c^\mathrm{ax}-c^\mathrm{ten}}3f^2
+
\frac{c^\mathrm{vec}-c^\mathrm{ten}}2\|v\|^2
+
c^\mathrm{ten}\overset{*}T{}_{\alpha\beta}\overset{*}T{}^{\alpha\beta}
-c^\mathrm{kin}\|\omega\|^2\right)\rho
\end{equation}
is the Lagrangian density.
Recall that the geometric quantities $f$, $v$, $\overset{*}T$ and $\omega$
appearing in formula
(\ref{Lagrangian of rotational elasticity equation 2})
are defined by formulae
(\ref{Simplifying the formula for potential energy equation 6}),
(\ref{Simplifying the formula for potential energy equation 7}),
(\ref{Measuring rotational deformations equation 12})
and
(\ref{Kinetic energy equation 3})
respectively.

Our construction of the action
(\ref{Lagrangian of rotational elasticity equation 1})
out of potential and kinetic energies
is Newtonian: compare with classical linear elasticity or even the harmonic oscillator
in classical mechanics.
An alternative approach is the relativistic one which boils down to rewriting
the formula for potential energy
in Lorentzian signature in dimension 1+3, with this ``extended''
potential energy becoming the action.
The Newtonian and relativistic approaches are different which can be seen,
for example,
from the fact that the relativistic approach always
imposes a unique velocity of wave propagation (speed of light)
whereas with the Newtonian approach one expects to get at least
two distinct wave velocities.

Starting with Einstein, most authors working in the subject of
teleparallelism adopt the relativistic approach. We shall, however,
stick with the Newtonian approach
(\ref{Lagrangian of rotational elasticity equation 1}).

\section{Reformulating the problem in the language of spinors}
\label{Reformulating the problem in the language of spinors}

Our field equations (Euler--Lagrange equations) are obtained by
varying the action
(\ref{Lagrangian of rotational elasticity equation 1})
with respect to the
coframe $\vartheta$ and density $\rho$. Varying with respect to the
density $\rho$ is easy: this gives the field equation
$
\frac{c^\mathrm{ax}-c^\mathrm{ten}}3f^2
+
\frac{c^\mathrm{vec}-c^\mathrm{ten}}2\|v\|^2
+
c^\mathrm{ten}\overset{*}T{}_{\alpha\beta}\overset{*}T{}^{\alpha\beta}
-c^\mathrm{kin}\|\omega\|^2=0
$
which is equivalent to $L(\vartheta,\rho)=0$. Varying with respect
to the coframe $\vartheta$ is more difficult because we have to
maintain the kinematic constraint (\ref{constraint for coframe}).

This technical difficulty can be overcome by switching to a
different dynamical variable. Namely, it is known \cite{1001.4726}
that in dimension~$3$ a coframe $\vartheta$ and a (positive) density
$\rho$ are equivalent to a 2-component complex-valued spinor field
$
\xi=\xi^a=
\begin{pmatrix}
\xi^1\\
\xi^2
\end{pmatrix}
$ modulo the sign of~$\xi$.
The explicit formulae establishing this equivalence are
\begin{equation}
\label{formula for density}
\rho=\bar\xi^{\dot a}\sigma_{0\dot ab}\xi^b,
\end{equation}
\begin{equation}
\label{formula for coframe elements 1 and 2}
(\vartheta^1+i\vartheta^2)_\alpha=\rho^{-1}
\epsilon^{\dot c\dot b}\sigma_{0\dot ba}\xi^a\sigma_{\alpha\dot cd}\xi^d,
\end{equation}
\begin{equation}
\label{formula for coframe element 3}
\vartheta^3{}_\alpha=\rho^{-1}
\bar\xi^{\dot a}\sigma_{\alpha\dot ab}\xi^b.
\end{equation}
Here $\sigma$ are Pauli matrices and $\epsilon$ is ``metric spinor''
(see (\ref{metric spinor})--(\ref{Pauli matrices spatial})),
the free tensor index $\alpha$ runs through the values
$1,2,3$, and the spinor summation indices run through the values
$1,2$ or $\dot1,\dot2$.
The advantage of switching to a spinor
field $\xi$ is that there are no kinematic constraints on its
components, so the derivation of field equations becomes
straightforward.

We give for reference more explicit versions of formulae
(\ref{formula for density})--(\ref{formula for coframe element 3}):
\begin{equation}
\label{formula for density explicit}
\rho=\bar\xi^{\dot 1}\xi^1+\bar\xi^{\dot 2}\xi^2,
\end{equation}
\begin{equation}
\label{formula for coframe elements 1 and 2 explicit}
(\vartheta^1+i\vartheta^2)_\alpha
=(\bar\xi^{\dot 1}\xi^1+\bar\xi^{\dot 2}\xi^2)^{-1}
\begin{pmatrix}
(\xi^1)^2-(\xi^2)^2\\
i(\xi^1)^2+i(\xi^2)^2\\
-2\xi^1\xi^2
\end{pmatrix},
\end{equation}
\begin{equation}
\label{formula for coframe element 3 explicit}
\vartheta^3{}_\alpha
=(\bar\xi^{\dot 1}\xi^1+\bar\xi^{\dot 2}\xi^2)^{-1}
\begin{pmatrix}
\bar\xi^{\dot 2}\xi^1+\bar\xi^{\dot 1}\xi^2\\
i\bar\xi^{\dot 2}\xi^1-i\bar\xi^{\dot 1}\xi^2\\
\bar\xi^{\dot 1}\xi^1-\bar\xi^{\dot 2}\xi^2
\end{pmatrix}.
\end{equation}

Let us rewrite the geometric quantities $f$, $v$, $\overset{*}T$ and
$\omega$ appearing in formula
(\ref{Lagrangian of rotational elasticity equation 2})
in terms of the spinor field $\xi$. The spinor representation of angular
velocity $\omega$ was derived in \cite{1001.4726}:
\begin{equation}
\label{angular velocity via spinor}
\omega_\alpha=i
\frac{
\bar\xi^{\dot a}\sigma_{\alpha\dot ab}\partial_0\xi^b
-
\xi^b\sigma_{\alpha\dot ab}\partial_0\bar\xi^{\dot a}
}
{
\bar\xi^{\dot c}\sigma_{0\dot cd}\xi^d
}
\end{equation}
or, more explicitly,
\begin{equation}
\label{angular velocity via spinor explicit}
\omega_\alpha=
\frac1{\bar\xi^{\dot 1}\xi^1+\bar\xi^{\dot 2}\xi^2}
\begin{pmatrix}
i\bar\xi^{\dot 2}\partial_0\xi^1+i\bar\xi^{\dot 1}\partial_0\xi^2\\
-\bar\xi^{\dot 2}\partial_0\xi^1+\bar\xi^{\dot 1}\partial_0\xi^2\\
i\bar\xi^{\dot 1}\partial_0\xi^1-i\bar\xi^{\dot 2}\partial_0\xi^2
\end{pmatrix}+\mathrm{c.c.}
\end{equation}
where the ``c.c.'' stands for ``complex conjugate term''.
The spinor representation of  the tensor $\overset{*}T$
is derived in
Appendix~\ref{Spinor representation of torsion},
see formula
(\ref{torsion via spinor})
or its more explicit version
(\ref{torsion via spinor explicit}).
Substituting (\ref{torsion via spinor})
or
(\ref{torsion via spinor explicit})
into
(\ref{Simplifying the formula for potential energy equation 1})
and
(\ref{Simplifying the formula for potential energy equation 2})
we arrive at spinor representations for the (pseudo)scalar $f$ and
vector $v$:
\begin{equation}
\label{Reformulating the problem in the language of spinors equation 1}
f=-2i
\frac{
\bar\xi^{\dot a}\sigma^\alpha{}_{\dot ab}\partial_\alpha\xi^b
-
\xi^b\sigma^\alpha{}_{\dot ab}\partial_\alpha\bar\xi^{\dot a}
}
{
\bar\xi^{\dot c}\sigma_{0\dot cd}\xi^d
}\,,
\end{equation}
\begin{equation}
\label{Reformulating the problem in the language of spinors equation 2}
v_\alpha=
-i\varepsilon_{\beta\gamma\alpha}
\frac{
\bar\xi^{\dot a}\sigma^\beta{}_{\dot ab}\partial^\gamma\xi^b
-
\xi^b\sigma^\beta{}_{\dot ab}\partial^\gamma\bar\xi^{\dot a}
}
{
\bar\xi^{\dot c}\sigma_{0\dot cd}\xi^d
}
\end{equation}
where $\partial^\alpha:=g^{\alpha\beta}\partial_\beta=\partial_\alpha$,
or, more explicitly,
\begin{equation}
\label{Reformulating the problem in the language of spinors equation 3}
f=
\frac2{\bar\xi^{\dot 1}\xi^1+\bar\xi^{\dot 2}\xi^2}
(
-i\bar\xi^{\dot 1}\partial_1\xi^2
-i\bar\xi^{\dot 2}\partial_1\xi^1
-\bar\xi^{\dot 1}\partial_2\xi^2
+\bar\xi^{\dot 2}\partial_2\xi^1
-i\bar\xi^{\dot 1}\partial_3\xi^1
+i\bar\xi^{\dot 2}\partial_3\xi^2
)+\mathrm{c.c.},
\end{equation}
\begin{equation}
\label{Reformulating the problem in the language of spinors equation 4}
v_\alpha=
\frac1{\bar\xi^{\dot 1}\xi^1+\bar\xi^{\dot 2}\xi^2}
\begin{pmatrix}
i\bar\xi^{\dot 1}\partial^2\xi^1
-i\bar\xi^{\dot 2}\partial^2\xi^2
-\bar\xi^{\dot 1}\partial^3\xi^2
+\bar\xi^{\dot 2}\partial^3\xi^1
\\
i\bar\xi^{\dot 1}\partial^3\xi^2
+i\bar\xi^{\dot 2}\partial^3\xi^1
-i\bar\xi^{\dot 1}\partial^1\xi^1
+i\bar\xi^{\dot 2}\partial^1\xi^2
\\
\bar\xi^{\dot 1}\partial^1\xi^2
-\bar\xi^{\dot 2}\partial^1\xi^1
-i\bar\xi^{\dot 1}\partial^2\xi^2
-i\bar\xi^{\dot 2}\partial^2\xi^1
\end{pmatrix}+\mathrm{c.c.}
\end{equation}
Note that formula
(\ref{Reformulating the problem in the language of spinors equation 1})
is a rephrasing of formula (B.5) from \cite{1001.4726}.

From now on we write our action
(\ref{Lagrangian of rotational elasticity equation 1})
and Lagrangian density
(\ref{Lagrangian of rotational elasticity equation 2})
as $S(\xi)$ and $L(\xi)$ rather than $S(\vartheta,\rho)$ and $L(\vartheta,\rho)$,
thus indicating that we have switched to spinors.
The explicit formula for $L(\xi)$ is obtained by substituting
formulae
(\ref{formula for density}),
(\ref{Reformulating the problem in the language of spinors equation 1}),
(\ref{Reformulating the problem in the language of spinors equation 2}),
(\ref{torsion via spinor}) and
(\ref{angular velocity via spinor})
into
(\ref{Lagrangian of rotational elasticity equation 2}).
The nonvanishing
spinor field $\xi$ is the new dynamical variable and it will be
varied without any constraints.

\section{Euler--Lagrange equation}
\label{Euler--Lagrange equation}

Let us perform a formal variation of our spinor field
$\xi\mapsto\xi+\delta\xi$, where
$\delta\xi:\mathbb{R}\times\mathbb{R}^3\to\mathbb{C}^2$
is an arbitrary (infinitely) smooth function with compact support.
Then, after integration by parts, the variation of our
action can be written as
\begin{equation}
\label{Euler--Lagrange equation equation 1}
\delta S
=\int
(F_{\dot a}\delta\bar\xi^{\dot a}+\bar F_a\delta\xi^a)
\,dx^0dx^1dx^2dx^3
\end{equation}
where $F$ is a dotted spinor field uniquely determined
by the undotted spinor field $\xi$. The Euler--Lagrange for
our unknown spinor field $\xi$ is, therefore,
\begin{equation}
\label{Euler--Lagrange equation equation 2}
F=0.
\end{equation}
The map
\begin{equation}
\label{Euler--Lagrange equation equation 3}
\xi\mapsto F
\end{equation}
defines a nonlinear second order partial
differential operator in the variables $x^0$ (time) and $x^\alpha$,
$\alpha=1,2,3$ (Cartesian coordinates).

We shall refrain from writing down the Euler--Lagrange equation
(\ref{Euler--Lagrange equation equation 2}) explicitly. The reason
for this is that in the current paper we are interested in finding a
particular class of solutions for which the procedure is much
simpler.

Note that for the special case of a purely axial material,
i.e.~material with
\begin{equation}
\label{purely axial material conditions 1}
c^{\mathrm{vec}}=0,
\qquad
c^{\mathrm{ten}}=0,
\end{equation}
the Euler--Lagrange equation (\ref{Euler--Lagrange equation equation 2}) was
written down explicitly in~\cite{1001.4726}. The calculations in
\cite{1001.4726} were carried out under the additional assumption
\begin{equation}
\label{purely axial material conditions 2}
c^{\mathrm{kin}}=\frac43c^{\mathrm{ax}}
\end{equation}
which can always be achieved by rescaling time $x^0$.

\section{Plane wave solutions}
\label{Plane wave solutions}

We seek solutions of the form
\begin{equation}
\label{Plane wave solutions equation 1}
\xi(\mathbf{x})=e^{-i\mathbf{p}\cdot\mathbf{x}}\zeta
\end{equation}
where $\zeta\ne0$ is a constant (complex) spinor and $\mathbf{p}$ is a
constant real covector. Here we use relativistic notation,
incorporating time $x^0$ into our coordinates. This means that
$\mathbf{x}=(x^0,x^1,x^2,x^3)$ and $\mathbf{p}=(p_0,p_1,p_2,p_3)$;
bold type indicates that we are working in (1+3)-dimensional spacetime.
The number $|p_0|$ is the wave frequency and the covector
$(p_1,p_2,p_3)$ is the wave vector in original 3-dimensional
Euclidean space. The 4-component covector $\mathbf{p}=(p_0,p_1,p_2,p_3)$
has the meaning of relativistic 4-momentum.

Throughout this section as well as the next one we assume that
\begin{equation}
\label{Plane wave solutions equation 2}
p_0\ne0
\end{equation}
which means that we are not interested in static (time-independent) solutions.
The sign of $p_0$ can be arbitrary.

Our Euler--Lagrange equation (\ref{Euler--Lagrange equation equation 2})
is highly nonlinear so it is by no means obvious that one can seek
solutions in the form of plane waves
(\ref{Plane wave solutions equation 1}).
Fortunately (and miraculously) this is the case.
In order to see this, we rewrite our
Euler--Lagrange equation (\ref{Euler--Lagrange equation equation 2})
in equivalent form
\begin{equation}
\label{Plane wave solutions equation 3}
e^{i\mathbf{p}\cdot\mathbf{x}}F=0.
\end{equation}
Note that the sign in the exponent in
(\ref{Plane wave solutions equation 3})
is opposite to that in (\ref{Plane wave solutions equation 1}).
We have

\begin{lemma}
\label{lemma 1}
If the spinor field $\xi$ is a plane wave
(\ref{Plane wave solutions equation 1})
then the left-hand side of equation
(\ref{Plane wave solutions equation 3}) is constant,
i.e.~it does not depend on $\mathbf{x}$.
\end{lemma}

Of course, Lemma \ref{lemma 1} can be equivalently reformulated as follows:
the nonlinear partial differential operator
(\ref{Euler--Lagrange equation equation 3})
maps a plane wave
(\ref{Plane wave solutions equation 1})
into a plane wave with the same relativistic 4-momentum
$\mathbf{p}$.

The proof of Lemma \ref{lemma 1} is quite technical and is given in
Appendix \ref{Separation of variables}.

Lemma \ref{lemma 1} justifies separation of variables, i.e.~it
reduces the study of the nonlinear partial differential equation
(\ref{Euler--Lagrange equation equation 2})
for the unknown spinor field $\xi$
to the study of the rational algebraic equation
(\ref{Plane wave solutions equation 3})
for the unknown constant spinor $\zeta$.
We suspect that the underlying group-theoretic reason for our
nonlinear partial differential equation
(\ref{Euler--Lagrange equation equation 2})
admitting separation of variables is the fact
that our model is $\mathrm{U}(1)$-invariant, i.e.~it is invariant
under the multiplication of the spinor field $\xi$ by a complex
constant of modulus 1. Hence, it is feasible that one could
prove Lemma \ref{lemma 1}, as well as Lemma \ref{lemma 2} stated
further down in this section, without performing the laborious
calculations presented in Appendix \ref{Separation of variables}.

We are now faced with the task of writing down the LHS of equation
(\ref{Plane wave solutions equation 3})
explicitly and with minimal
calculations. To this end we address a seemingly different issue: we
examine what happens when we substitute our plane wave
(\ref{Plane wave solutions equation 1}) into our Lagrangian density
$L(\xi)$, rather than the Euler--Lagrange equation
(\ref{Euler--Lagrange equation equation 2}).

Substituting formula
(\ref{Plane wave solutions equation 1})
into formulae
(\ref{formula for density}),
(\ref{Reformulating the problem in the language of spinors equation 1}),
(\ref{Reformulating the problem in the language of spinors equation 2}),
(\ref{torsion via spinor}) and
(\ref{angular velocity via spinor})
we get
\begin{equation}
\label{Plane wave solutions equation 4}
\rho=\bar\zeta^{\dot a}\sigma_{0\dot ab}\zeta^b,
\end{equation}
\begin{equation}
\label{Plane wave solutions equation 5}
f=-
\frac{4
\bar\zeta^{\dot a}\sigma^\alpha{}_{\dot ab}p_\alpha\zeta^b
}
{
\bar\zeta^{\dot c}\sigma_{0\dot cd}\zeta^d
}\,,
\end{equation}
\begin{equation}
\label{Plane wave solutions equation 6}
v_\alpha=
-\frac{2
\varepsilon_{\beta\gamma\alpha}
\bar\zeta^{\dot a}\sigma^\beta{}_{\dot ab}p^\gamma\zeta^b
}
{
\bar\zeta^{\dot c}\sigma_{0\dot cd}\zeta^d
}\,,
\end{equation}
\begin{equation}
\label{Plane wave solutions equation 7}
\overset{*}T_{\alpha\beta}=2
\frac{
\bar\zeta^{\dot a}\sigma_{\beta\dot ab}p_\alpha\zeta^b
-
\bar\zeta^{\dot a}\sigma^\gamma{}_{\dot ab}p_\gamma\zeta^b
g_{\alpha\beta}
}
{
\bar\zeta^{\dot c}\sigma_{0\dot cd}\zeta^d
}\,,
\end{equation}
\begin{equation}
\label{Plane wave solutions equation 8}
\omega_\alpha=
\frac{2
\bar\zeta^{\dot a}\sigma_{\alpha\dot ab}p_0\zeta^b
}
{
\bar\zeta^{\dot c}\sigma_{0\dot cd}\zeta^d
}\,,
\end{equation}
or, more explicitly,
\begin{equation}
\label{Plane wave solutions equation 9}
\rho=\bar\zeta^{\dot 1}\zeta^1+\bar\zeta^{\dot 2}\zeta^2,
\end{equation}
\begin{equation}
\label{Plane wave solutions equation 10}
f=
\frac{4}{\bar\zeta^{\dot 1}\zeta^1+\bar\zeta^{\dot 2}\zeta^2}
\left(
 p_1(-\bar\zeta^{\dot 1}\zeta^2-\bar\zeta^{\dot 2}\zeta^1)
+ip_2(\bar\zeta^{\dot 1}\zeta^2-\bar\zeta^{\dot 2}\zeta^1)
+p_3(-\bar\zeta^{\dot 1}\zeta^1+\bar\zeta^{\dot 2}\zeta^2)
\right),
\end{equation}
\begin{equation}
\label{Plane wave solutions equation 11}
v_\alpha=
\frac{2}{\bar{\zeta}^{\dot 1}\zeta^1+\bar{\zeta}^{\dot 2}\zeta^2}
\begin{pmatrix}
p_2\left(\bar\zeta^{\dot 1}\zeta^1-\bar\zeta^{\dot 2}\zeta^2\right)
+ip_3\left(\bar\zeta^{\dot 1}\zeta^2-\bar\zeta^{\dot 2}\zeta^1\right)
\\
p_3\left(\bar\zeta^{\dot 1}\zeta^2+\bar\zeta^{\dot 2}\zeta^1\right)
+p_1\left(-\bar\zeta^{\dot 1}\zeta^1+\bar\zeta^{\dot 2}\zeta^2\right)
\\
ip_1\left(-\bar\zeta^{\dot 1}\zeta^2+\bar\zeta^{\dot 2}\zeta^1\right)
+p_2\left(-\bar\zeta^{\dot 1}\zeta^2-\bar\zeta^{\dot 2}\zeta^1\right)
\end{pmatrix},
\end{equation}
\begin{equation}
\label{Plane wave solutions equation 12}
\begin{pmatrix}
\overset{*}T_{11}\\
\overset{*}T_{12}\\
\overset{*}T_{13}\\
\overset{*}T_{21}\\
\overset{*}T_{22}\\
\overset{*}T_{23}\\
\overset{*}T_{31}\\
\overset{*}T_{32}\\
\overset{*}T_{33}
\end{pmatrix}
=
\frac{2}{\bar\zeta^{\dot 1}\zeta^1+\bar\zeta^{\dot 2}\zeta^2}
\begin{pmatrix}
ip_2 \left(\bar\zeta^{\dot 1}\zeta^2-\bar\zeta^{\dot 2}\zeta^1\right)
+p_3 \left(-\bar\zeta^{\dot 1}\zeta^1+\bar\zeta^{\dot 2}\zeta^2\right)
\\
ip_1 \left(-\bar{\zeta}^{\dot 1}\zeta^2+\bar{\zeta}^{\dot 2}\zeta^1\right)
\\
p_1 \left(\bar{\zeta}^{\dot 1}\zeta^1-\bar{\zeta}^{\dot 2}\zeta^2\right)
\\
p_2 \left(\bar{\zeta}^{\dot 1}\zeta^2+\bar{\zeta}^{\dot 2}\zeta^1\right)
\\
p_3 \left(-\bar\zeta^{\dot 1}\zeta^1+\bar\zeta^{\dot 2}\zeta^2\right)
+p_1 \left(-\bar\zeta^{\dot 1}\zeta^2-\bar\zeta^{\dot 2}\zeta^1\right)
\\
p_2 \left(\bar\zeta^{\dot 1}\zeta^1-\bar\zeta^{\dot 2}\zeta^2\right)
\\
p_3 \left(\bar\zeta^{\dot 1}\zeta^2+\bar\zeta^{\dot 2}\zeta^1\right)
\\
ip_3 \left(-\bar\zeta^{\dot 1}\zeta^2+\bar\zeta^{\dot 2}\zeta^1\right)
\\
p_1 \left(-\bar{\zeta}^{\dot 1}\zeta^2-\bar{\zeta}^{\dot 2}\zeta^1\right)
+ip_2 \left(\bar{\zeta}^{\dot 1}\zeta^2-\bar{\zeta}^{\dot 2}\zeta^1\right)
\end{pmatrix},
\end{equation}
\begin{equation}
\label{Plane wave solutions equation 13}
\omega_\alpha=
\frac{2p_0}{\bar\zeta^{\dot 1}\zeta^1+\bar\zeta^{\dot 2}\zeta^2}
\begin{pmatrix}
-i\bar\zeta^{\dot 1}\zeta^2-i\bar\zeta^{\dot 2}\zeta^1\\
-\bar\zeta^{\dot 1}\zeta^2+\bar\zeta^{\dot 2}\zeta^1\\
-i\bar\zeta^{\dot 1}\zeta^1+i\bar\zeta^{\dot 2}\zeta^2
\end{pmatrix}.
\end{equation}

Substituting formulae
(\ref{Plane wave solutions equation 4})--(\ref{Plane wave solutions equation 8})
or their more explicit versions
(\ref{Plane wave solutions equation 9})--(\ref{Plane wave solutions equation 13})
into formula (\ref{Lagrangian of rotational elasticity equation 2})
we arrive at a Lagrangian density
$L(\zeta;\mathbf{p})$ which does not depend on $\mathbf{x}$.
The self-contained formula for this
Lagrangian density, written in terms of 4-momentum
$\mathbf{p}$ and 4-current
\begin{equation}
\label{Plane wave solutions equation 14}
\mathbf{j}_{\bm{\alpha}}:=\bar\zeta^{\dot a}\sigma_{\bm{\alpha}\dot ab}\zeta^b,
\end{equation}
is
\begin{equation}
\label{Plane wave solutions equation 15}
L(\zeta;\mathbf{p})\!=\!
\frac{2}{j_0}\left(
c^{\mathrm{vec}}\!+c^{\mathrm{ten}}
\right)\|p\|^2\|j\|^2
+\frac{4}{j_0}\left(
\frac43c^{\mathrm{ax}}\!-\frac12c^{\mathrm{vec}}\!+\frac16c^{\mathrm{ten}}
\right)(j\cdot p)^2
-\frac{4}{j_0}c^{\mathrm{kin}}p_0^2\|j\|^2\!.
\end{equation}
Here we write our 4-covectors as
$\mathbf{p}=(p_0,p)$ and $\mathbf{j}=(j_0,j)$,
where $p$ and $j$ are 3-covectors.

We view the 4-momentum $\mathbf{p}$ as a parameter
and the constant spinor $\zeta\ne0$ as the dynamical variable.
The Lagrangian density $L(\zeta;\mathbf{p})$ is a smooth function
of $\operatorname{Re}\zeta$ and $\operatorname{Im}\zeta$,
so varying $\zeta$ we get
\begin{equation}
\label{Plane wave solutions equation 16}
\delta L
=G_{\dot a}\delta\bar\zeta^{\dot a}+\bar G_a\delta\zeta^a
\end{equation}
where $G$ is a dotted constant spinor expressed via the partial
derivatives of $L(\zeta;\mathbf{p})$ with respect to
$\operatorname{Re}\zeta$ and $\operatorname{Im}\zeta$.
It is natural to ask the question: what is the relationship between
the spinor field $F$ appearing in formula
(\ref{Euler--Lagrange equation equation 1})
and the constant spinor
$G$ appearing in formula
(\ref{Plane wave solutions equation 16})?
The answer is given by

\begin{lemma}
\label{lemma 2}
If the spinor field $\xi$ is a plane wave
(\ref{Plane wave solutions equation 1})
then $G=e^{i\mathbf{p}\cdot\mathbf{x}}F$.
\end{lemma}

The proof of Lemma \ref{lemma 2} is presented in Appendix
\ref{Separation of variables}.

Lemma \ref{lemma 2} reduces the construction of plane wave solutions
of rotational elasticity to finding the critical, with respect to
$\zeta$, points of the function $L(\zeta;\mathbf{p})$.
Varying (\ref{Plane wave solutions equation 15}), we arrive
at the following equation for critical points:
\begin{multline}
\label{Plane wave solutions equation 17}
\frac{4}{j_0}\left(
c^{\mathrm{vec}}+c^{\mathrm{ten}}
\right)\|p\|^2j^\alpha\sigma_{\alpha\dot ab}\zeta^b
-\frac{2}{j_0^2}\left(
c^{\mathrm{vec}}+c^{\mathrm{ten}}
\right)\|p\|^2\|j\|^2\sigma_{0\dot ab}\zeta^b
\\
+\frac{8}{j_0}\left(
\frac43c^{\mathrm{ax}}-\frac12c^{\mathrm{vec}}+\frac16c^{\mathrm{ten}}
\right)(j\cdot p)p^\alpha\sigma_{\alpha\dot ab}\zeta^b
-\frac{4}{j_0^2}\left(
\frac43c^{\mathrm{ax}}-\frac12c^{\mathrm{vec}}+\frac16c^{\mathrm{ten}}
\right)(j\cdot p)^2\sigma_{0\dot ab}\zeta^b
\\
-\frac{8}{j_0}c^{\mathrm{kin}}p_0^2j^\alpha\sigma_{\alpha\dot ab}\zeta^b
+\frac4{j_0^2}c^{\mathrm{kin}}{p_0}^2\|j\|^2\sigma_{0\dot ab}\zeta^b=0.
\end{multline}
Recall that the 4-current $\mathbf{j}=(j_0,j)$ appearing in the
above equation is defined in accordance with formula
(\ref{Plane wave solutions equation 14}).

It now remains to find the 4-momenta $\mathbf{p}$
and spinors $\zeta\ne0$ which satisfy equation
(\ref{Plane wave solutions equation 17}).
We carry out the analysis of equation
(\ref{Plane wave solutions equation 17})
assuming that
\begin{equation}
\label{Plane wave solutions equation 18}
\rho=j_0=\bar\zeta^{\dot 1}\zeta^1+\bar\zeta^{\dot 2}\zeta^2=1.
\end{equation}
Condition (\ref{Plane wave solutions equation 18}) is a
normalisation of the density: general plane wave solutions
differ from those satisfying condition (\ref{Plane wave solutions equation 18})
by a real scaling factor.
Furthermore, we assume that
\begin{equation}
\label{Plane wave solutions equation 19}
\zeta^b=
\begin{pmatrix}1\\0\end{pmatrix}.
\end{equation}
Indeed, any spinor $\zeta$ satisfying
condition (\ref{Plane wave solutions equation 18})
can be transformed into the spinor
(\ref{Plane wave solutions equation 19})
by the linear transformation
\begin{equation}
\label{Plane wave solutions equation 20}
\zeta\mapsto U\zeta
\end{equation}
where $U$ is a special ($\det U=1$) unitary matrix.
The transformation (\ref{Plane wave solutions equation 20})
leads to a rotation of the spatial part of the 4-current
(\ref{Plane wave solutions equation 14}), so plane wave solutions
with general $\zeta$ differ from those with $\zeta$ of the form
(\ref{Plane wave solutions equation 19}) by a rotation.

Substituting
(\ref{Plane wave solutions equation 17}),
(\ref{Plane wave solutions equation 19}),
(\ref{Pauli matrix temporal}),
(\ref{Pauli matrices spatial})
and
(\ref{Plane wave solutions equation 13})
into
(\ref{Plane wave solutions equation 16})
we get
\begin{multline*}
4\left(
c^{\mathrm{vec}}+c^{\mathrm{ten}}
\right)\|p\|^2\begin{pmatrix}1\\0\end{pmatrix}
-2\left(
c^{\mathrm{vec}}+c^{\mathrm{ten}}
\right)\|p\|^2\begin{pmatrix}1\\0\end{pmatrix}
\\
+8\left(
\frac43c^{\mathrm{ax}}-\frac12c^{\mathrm{vec}}+\frac16c^{\mathrm{ten}}
\right)
p_3\begin{pmatrix}p_3\\ p_1+ip_2\end{pmatrix}
-4\left(
\frac43c^{\mathrm{ax}}-\frac12c^{\mathrm{vec}}+\frac16c^{\mathrm{ten}}
\right)p_3^2\begin{pmatrix}1\\0\end{pmatrix}
\\
-4c^{\mathrm{kin}}\begin{pmatrix}p_0^2\\0\end{pmatrix}=0,
\end{multline*}
or, equivalently,
\begin{equation}
\label{Plane wave solutions equation 21}
2\left(
c^{\mathrm{vec}}+c^{\mathrm{ten}}
\right)\begin{pmatrix}\|p\|^2\\0\end{pmatrix}
\!+4\left(
\frac43c^{\mathrm{ax}}-\frac12c^{\mathrm{vec}}+\frac16c^{\mathrm{ten}}
\right)
\begin{pmatrix}p_3^2\\ 2p_3(p_1+ip_2)\end{pmatrix}
\!-4c^{\mathrm{kin}}\begin{pmatrix}p_0^2\\0\end{pmatrix}
\!=0.
\end{equation}

Put
\begin{equation}
\label{Plane wave solutions equation 22}
v_1:=
\sqrt{\frac{4c^{\mathrm{ax}}+2c^{\mathrm{ten}}}{3c^{\mathrm{kin}}}}\ ,
\qquad
v_2:=
\sqrt{\frac{c^{\mathrm{vec}}+c^{\mathrm{ten}}}{2c^{\mathrm{kin}}}}\ .
\end{equation}
Note that because we assumed our three elastic moduli to be
nonnegative and not all zero, our $v_1$ and $v_2$ are nonnegative
and not both zero.
Using
(\ref{Plane wave solutions equation 22})
we can now rewrite
equation (\ref{Plane wave solutions equation 21})
in more compact form
\begin{equation}
\label{Plane wave solutions equation 23}
v_2^2
\begin{pmatrix}\|p\|^2\\0\end{pmatrix}
+(v_1^2-v_2^2)
\begin{pmatrix}p_3^2\\ 2p_3(p_1+ip_2)\end{pmatrix}
-\begin{pmatrix}p_0^2\\0\end{pmatrix}
=0.
\end{equation}

The analysis of equation (\ref{Plane wave solutions equation 23}) is
elementary and the outcome is summarised in the following theorem,
which is the main result of our paper.

\begin{Theorem}
\label{Theorem 1}
Plane wave solutions of rotational elasticity can,
up to rescaling and rotation, be explicitly written down in the form
(\ref{Plane wave solutions equation 1}),
(\ref{Plane wave solutions equation 19})
with arbitrary nonzero $p_0$ and $p=(p_1,p_2,p_3)$ determined as follows.
\begin{itemize}
\item
\ If $v_1>0$ and $v_2>0$ and $v_1\ne v_2$ then we have two possibilities:
\begin{itemize}
\item
$\displaystyle p=\left(0,0,\pm\frac{p_0}{v_1}\right)$ (type 1 wave), or
\item
$\displaystyle p=\left(
\frac{|p_0|}{v_2}\cos\varphi,\frac{|p_0|}{v_2}\sin\varphi,0\right)$
where $\varphi\in\mathbb{R}$ is arbitrary (type 2 wave).
\end{itemize}
\item
\ If $v_1>0$ and $v_2>0$ and $v_1=v_2$ then
$p$ is an arbitrary 3-vector satisfying
$\displaystyle\|p\|=\frac{|p_0|}{v_1}\,$.
\item
\ If $v_1>0$ and $v_2=0$ then
$\displaystyle p=\left(0,0,\pm\frac{p_0}{v_1}\right)$.
\item
\ If $v_1=0$ and $v_2>0$ then
$\displaystyle p=\left(
\frac{|p_0|}{v_2}\cos\varphi,\frac{|p_0|}{v_2}\sin\varphi,0\right)$
where $\varphi\in\mathbb{R}$ is arbitrary.
\end{itemize}
\end{Theorem}

\pagebreak

Theorem \ref{Theorem 1} shows that rotational elasticity, like
classical linear elasticity, produces two distinct types of plane
wave solutions. We call these solutions type 1 and type 2 and they
propagate with velocities $v_1$ and $v_2$ respectively, with $v_1$
and $v_2$ given by formulae (\ref{Plane wave solutions equation 22}).

However, unlike with classical linear elasticity, in rotational
elasticity the two wave velocities, $v_1$ and $v_2$, are not
ordered, i.e.~we do not know a priori which one, $v_1$ or $v_2$, is
bigger. The reason the two wave velocities are not ordered is
because rotational elasticity has three elastic moduli compared to
the two elastic moduli of classical linear elasticity. The ``extra''
elastic modulus is $c^{\mathrm{vec}}$, the one associated with the
antisymmetric part of the rank two tensor $\overset{*}T$ which we
use as a measure of rotational deformations. If we set
$c^{\mathrm{vec}}=0$, we end up with the inequality
$v_1\ge\sqrt{\frac43}\,v_2$ similar to the well known inequality
from classical linear elasticity, see formula (22.5) in \cite{MR884707}.

\section{The massless Dirac equation}
\label{The massless Dirac equation}

In this section we consider a purely axial material
(\ref{purely axial material conditions 1}),
(\ref{purely axial material conditions 2}).
Substituting
(\ref{purely axial material conditions 1})
and
(\ref{purely axial material conditions 2})
into
(\ref{Plane wave solutions equation 22})
we get $v_1=1$ and $v_2=0$,
so a purely axial material supports only type 1 waves.
Throughout this section we also retain the assumption
(\ref{Plane wave solutions equation 2}).

Note that a purely axial material has the remarkable property that
its potential energy is invariant under conformal rescalings of the
spatial metric by an arbitrary positive scalar function. We do not
elaborate on this issue in the current paper because we chose to
work with a specific (standard Euclidean) metric. The appropriate
arguments are presented in Section 2 of \cite{1001.4726}.

Our aim is to compare our model with the linear partial differential
equation (or, more precisely, system of two linear partial differential
equations)
\begin{equation}
\label{Weyl equation}
i(\pm\sigma^0{}_{\dot ab}\partial_0
+\sigma^\alpha{}_{\dot ab}\partial_\alpha)\xi^b=0.
\end{equation}
Here $\sigma$ are Pauli matrices
(\ref{Pauli matrix temporal}),
(\ref{Pauli matrices spatial}),
the free spinor index $\dot a$ runs through the values $\dot1,\dot2$,
summation is carried out over the tensor index $\alpha=1,2,3$
as well as the spinor index $b=1,2$,
and $\xi$ is the unknown spinor field.
We give for reference a more explicit version of equation
(\ref{Weyl equation}):
\begin{equation}
\label{Weyl equation explicit}
i
\begin{pmatrix}
\mp\partial_0+\partial_3&\partial_1-i\partial_2\\
\partial_1+i\partial_2&\mp\partial_0-\partial_3
\end{pmatrix}
\begin{pmatrix}
\xi^1\\
\xi^2
\end{pmatrix}
=0.
\end{equation}

Equation (\ref{Weyl equation}) is called \emph{the massless Dirac
equation} or \emph{the Weyl equation} This equation is the accepted
mathematical model for a massless neutrino field. The two choices of
sign in (\ref{Weyl equation}) give two versions of the Weyl equation
which differ by time reversal. Thus, we have a pair of Weyl
equations.

We want to compare plane wave solutions
(see (\ref{Plane wave solutions equation 1}))
of our model with those of the Weyl equation.
As both models are invariant under the rescaling of the spinor field
by a positive real constant
as well as the rotations of Euclidean 3-space, it is sufficient to
compare plane wave solutions for
$\zeta$ of the form
(\ref{Plane wave solutions equation 19}).
Substituting
(\ref{Plane wave solutions equation 1})
and
(\ref{Plane wave solutions equation 19})
into
(\ref{Weyl equation})
or its more explicit version
(\ref{Weyl equation explicit}) we get
$p=(0,0,\pm p_0)$ which is exactly what Theorem~\ref{Theorem 1} gives us.
Thus, we have established

\begin{Theorem}
\label{Theorem 2}
In the case of a purely axial material a plane wave spinor field is
a solution of rotational elasticity if and only if it is a solution
of one of the two Weyl equations~(\ref{Weyl equation}).
\end{Theorem}

It turns out that, in fact, a much stronger result
holds. Consider a spinor field of the form
\begin{equation}
\label{stationary spinor field}
\xi(x^0,x^1,x^2,x^3)=
e^{-ip_0x^0}\eta(x^1,x^2,x^3).
\end{equation}
We will call spinor fields of the form
(\ref{stationary spinor field}) \emph{stationary}.
In considering stationary spinor fields what we are doing
is separating out time only as opposed to separating out all
the variables.

The following result generalises Theorem~\ref{Theorem 2}.

\begin{Theorem}
\label{Theorem 3}
In the case of a purely axial material a nonvanishing stationary spinor field is
a solution of rotational elasticity if and only if it is a solution
of one of the two Weyl equations~(\ref{Weyl equation}).
\end{Theorem}

Theorem~\ref{Theorem 3} was proved in \cite{1001.4726} and the proof
is quite delicate. It involves an argument which reduces a nonlinear
second order partial differential equation of a particular type to a
pair of linear first order partial differential equations, which is,
effectively, a form of integrability. An abstract self-contained
version of this argument is given in Appendix B of~\cite{1007.3481}.

\section{Acknowledgments}

The authors are grateful to
J.~Burnett,
O.~Chervova,
F.~W.~Hehl
and Yu.~N.~Obukhov for
stimulating discussions.

\appendix

\section{Notation}
\label{Notation}

Our notation follows
\cite{MR2573111,1001.4726,1007.3481,prd2007}.
The only difference with
\cite{MR2573111,prd2007}
is that in the latter the spacetime metric
has opposite signature.
In \cite{1001.4726,1007.3481} the signature is the same as in the current paper,
i.e.~the 3-dimensional spatial metric
has signature $\,{+++}\,$.

We use Greek letters for tensor (holonomic) indices and Latin
letters for frame (anholonomic) indices. We identify differential
forms with covariant antisymmetric tensors.

We define the action of the Hodge star on a rank
$r$ antisymmetric tensor $R$ as
\begin{equation}
\label{definition of the Hodge star}
(*R)_{\alpha_{r+1}\ldots\alpha_3}:=(r!)^{-1}\,
R^{\alpha_1\ldots\alpha_r}\varepsilon_{\alpha_1\ldots\alpha_3}
\end{equation}
where $\varepsilon$ is the totally antisymmetric quantity,
$\varepsilon_{123}:=+1$.

We use two-component complex-valued spinors (Weyl spinors) whose
indices run through the values $1,2$ or $\dot1,\dot2$. Complex
conjugation makes the undotted indices dotted and vice versa.

We define the ``metric spinor''
\begin{equation}
\label{metric spinor}
\epsilon_{ab}=\epsilon_{\dot a\dot b}=
\epsilon^{ab}=\epsilon^{\dot a\dot b}=
\begin{pmatrix}
0&-1\\
1&0
\end{pmatrix}
\end{equation}
and choose Pauli matrices
\begin{equation}
\label{Pauli matrix temporal}
\sigma_{0\dot ab}=
\begin{pmatrix}
1&0\\
0&1
\end{pmatrix}
=-\sigma^0{}_{\dot ab},
\end{equation}
\begin{equation}
\label{Pauli matrices spatial}
\sigma_{1\dot ab}=
\begin{pmatrix}
0&1\\
1&0
\end{pmatrix}
=\sigma^1{}_{\dot ab},
\qquad
\sigma_{2\dot ab}=
\begin{pmatrix}
0&-i\\
i&0
\end{pmatrix}
=\sigma^2{}_{\dot ab},
\qquad
\sigma_{3\dot ab}=
\begin{pmatrix}
1&0\\
0&-1
\end{pmatrix}
=\sigma^3{}_{\dot ab}\,.
\end{equation}
Here the first index enumerates rows and the second enumerates
columns

\section{Spinor representation of torsion}
\label{Spinor representation of torsion}

We show in this appendix that the tensor
$\overset{*}T_{\alpha\beta}$ defined by formula
(\ref{Measuring rotational deformations equation 11})
(it is the Hodge dual, in the last pair indices, of the torsion tensor)
is expressed via the spinor field $\xi$ as
\begin{equation}
\label{torsion via spinor}
\overset{*}T_{\alpha\beta}=i
\frac{
\bar\xi^{\dot a}\sigma_{\beta\dot ab}\partial_\alpha\xi^b
-
\xi^b\sigma_{\beta\dot ab}\partial_\alpha\bar\xi^{\dot a}
-
(
\bar\xi^{\dot a}\sigma^\gamma{}_{\dot ab}\partial_\gamma\xi^b
-
\xi^b\sigma^\gamma{}_{\dot ab}\partial_\gamma\bar\xi^{\dot a}
)g_{\alpha\beta}
}
{
\bar\xi^{\dot c}\sigma_{0\dot cd}\xi^d
}\,.
\end{equation}
Note that formula (\ref{torsion via spinor})
is invariant under the rescaling of our spinor field by an arbitrary
positive scalar function.

Formula (\ref{torsion via spinor})
is proved by direct substitution of
formulae
(\ref{formula for coframe elements 1 and 2})
and
(\ref{formula for coframe element 3})
into
(\ref{Measuring rotational deformations equation 11}).
In order to simplify calculations we observe
that the expressions in the left- and right-hand sides of formula
(\ref{torsion via spinor})
have an invariant nature, hence it is sufficient to prove
formula (\ref{torsion via spinor})
at a point at which the spinor field takes
the value
$\xi^a=
\begin{pmatrix}
1\\
0
\end{pmatrix}$.
Then at this point we have
\begin{equation}
\label{standard coframe}
\vartheta^j{}_\beta=\delta^j{}_\beta\,,
\end{equation}
\begin{equation}
\label{partial derivative of coframe}
[\partial_{\alpha}(\vartheta^1+i\vartheta^2)]_\beta=
\begin{pmatrix}
\partial_{\alpha}\xi^1-\partial_{\alpha}\bar\xi^{\dot1}\\
i\partial_{\alpha}\xi^1-i\partial_{\alpha}\bar\xi^{\dot1}\\
-2\partial_{\alpha}\xi^2
\end{pmatrix},
\qquad
[\partial_{\alpha}\vartheta^3]_\beta=
\begin{pmatrix}
\partial_{\alpha}\xi^2+\partial_{\alpha}\bar\xi^{\dot2}\\
-i\partial_{\alpha}\xi^2+i\partial_{\alpha}\bar\xi^{\dot2}\\
0
\end{pmatrix},
\end{equation}
where $\alpha=1,2,3$.
Note that formulae (\ref{partial derivative of coframe}) imply
\begin{equation}
\label{curl of vartheta1 plus ivartheta2}
[\operatorname{curl}(\vartheta^1+i\vartheta^2)]_\beta=
\begin{pmatrix}
-2\partial_2\xi^2-\partial_3(i\xi^1-i\bar\xi^{\dot1})\\
2\partial_1\xi^2+\partial_3(\xi^1-\bar\xi^{\dot1})\\
\partial_1(i\xi^1-i\bar\xi^{\dot1})-\partial_2(\xi^1-\bar\xi^{\dot1})
\end{pmatrix},
\end{equation}
\begin{equation}
\label{curl of vartheta3}
[\operatorname{curl}\vartheta^3]_\beta=
\begin{pmatrix}
-\partial_3(-i\xi^2+i\bar\xi^{\dot2})\\
\partial_3(\xi^2+\bar\xi^{\dot2})\\
\partial_1(-i\xi^2+i\bar\xi^{\dot2})-\partial_2(\xi^2+\bar\xi^{\dot2})
\end{pmatrix}.
\end{equation}

We now rewrite formula
(\ref{Measuring rotational deformations equation 11})
in the form
\begin{equation}
\label{convenient formula for T star}
\overset{*}T
=\frac12(\vartheta^1-i\vartheta^2)
\otimes\operatorname{curl}(\vartheta^1+i\vartheta^2)
+\frac12(\vartheta^1+i\vartheta^2)
\otimes\operatorname{curl}(\vartheta^1-i\vartheta^2)
+\vartheta^3\otimes\operatorname{curl}\vartheta^3.
\end{equation}
Substituting formulae (\ref{standard coframe}),
(\ref{curl of vartheta1 plus ivartheta2})
and
(\ref{curl of vartheta3})
into formula (\ref{convenient formula for T star})
we get
\[
\overset{*}T_{\alpha\beta}=
\begin{pmatrix}
-\partial_2(\xi^2+\bar\xi^{\dot2})-i\partial_3(\xi^1-\bar\xi^{\dot1})&
\partial_1(\xi^2+\bar\xi^{\dot2})&
i\partial_1(\xi^1-\bar\xi^{\dot1})\\
i\partial_2(\xi^2-\bar\xi^{\dot2})&
-i\partial_1(\xi^2-\bar\xi^{\dot2})-i\partial_3(\xi^1-\bar\xi^{\dot1})&
i\partial_2(\xi^1-\bar\xi^{\dot1})\\
i\partial_3(\xi^2-\bar\xi^{\dot2})&
\partial_3(\xi^2+\bar\xi^{\dot2})&
-i\partial_1(\xi^2-\bar\xi^{\dot2})-\partial_2(\xi^2+\bar\xi^{\dot2})
\end{pmatrix}
\]
which coincides with the RHS of formula
(\ref{torsion via spinor}). This completes the proof.

We give for reference a more explicit version of formula
(\ref{torsion via spinor}):
\begin{equation}
\label{torsion via spinor explicit}
\begin{pmatrix}
\overset{*}T_{11}\\
\overset{*}T_{12}\\
\overset{*}T_{13}\\
\overset{*}T_{21}\\
\overset{*}T_{22}\\
\overset{*}T_{23}\\
\overset{*}T_{31}\\
\overset{*}T_{32}\\
\overset{*}T_{33}
\end{pmatrix}=
\frac1{\bar\xi^{\dot 1}\xi^1+\bar\xi^{\dot 2}\xi^2}
\begin{pmatrix}
\bar\xi^{\dot 2}\partial_2\xi^1
-\bar\xi^{\dot 1}\partial_2\xi^2
+i\bar\xi^{\dot 2}\partial_3\xi^2
-i\bar\xi^{\dot 1}\partial_3\xi^1
\\
\bar\xi^{\dot 1}\partial_1\xi^2
-\bar{\xi}^{\dot 2}\partial_1\xi^1
\\
i\bar\xi^{\dot 1}\partial_1\xi^1
-i\bar{\xi}^{\dot 2}\partial_1\xi^2
\\
i\bar{\xi}^{\dot 2}\partial_2\xi^1
+i\bar{\xi}^{\dot 1}\partial_2\xi^2
\\
-i\bar\xi^{\dot 1}\partial_1\xi^2
-i\bar\xi^{\dot 2}\partial_1\xi^1
-i\bar\xi^{\dot 1}\partial_3\xi^1
+i\bar\xi^{\dot 2}\partial_3\xi^2
\\
i\bar\xi^{\dot 1}\partial_2\xi^1
-i\bar\xi^{\dot 2}\partial_2\xi^2
\\
i\bar\xi^{\dot 1}\partial_3\xi^2
+i\bar\xi^{\dot 2}\partial_3\xi^1
\\
\bar\xi^{\dot 1}\partial_3\xi^2
-\bar\xi^{\dot 2}\partial_3\xi^1
\\
-i\bar{\xi}^{\dot 1}\partial_1\xi^2
-i\bar{\xi}^{\dot 2}\partial_1\xi^1
+\bar{\xi}^{\dot 2}\partial_2\xi^1
-\bar{\xi}^{\dot 1}\partial_2\xi^2
\end{pmatrix}+\mathrm{c.c.}
\end{equation}

\pagebreak

\section{Separation of variables}
\label{Separation of variables}

In this appendix we prove Lemmata
\ref{lemma 1} and \ref{lemma 2}.
Note that it would suffice to prove Lemma~\ref{lemma 2} only, because
Lemma~\ref{lemma 1} follows from Lemma~\ref{lemma 2}.
However, we prove Lemma~\ref{lemma 1} first for the sake of
clarity of exposition.

Let us arrange the (pseudo)scalar $f$, the three components of the
vector $v_\alpha$, the nine components of the tensor
$\overset{*}T_{\alpha\beta}$ and the three components of the
(pseudo)vector $\omega_\alpha$ into one 16-component ``vector''
$V_J$, $J=1,\ldots,16$. Then our Lagrangian density
(\ref{Lagrangian of rotational elasticity equation 2})
can be written as
\begin{equation}
\label{Separation of variables equation 1}
L(\xi)=\rho\sum_{J=1}^{16}A_JV_J^2
\end{equation}
where the $A_J$ are some real constants.
Put $W_J:=\sqrt\rho\,V_J$. Then formula
(\ref{Separation of variables equation 1}) takes the form
\begin{equation}
\label{Separation of variables equation 2}
L(\xi)=\sum_{J=1}^{16}A_JW_J^2.
\end{equation}
According to formulae
(\ref{Reformulating the problem in the language of spinors equation 1}),
(\ref{Reformulating the problem in the language of spinors equation 2}),
(\ref{torsion via spinor}),
(\ref{angular velocity via spinor}) and
(\ref{formula for density})
the components of the ``vector'' $W$ are expressed via the
spinor field $\xi$ as
\begin{equation}
\label{Separation of variables equation 3}
W_J=
i\frac{
\bar\xi^{\dot a}B_J^{\bm{\alpha}}{}_{\dot ab}
\partial_{\bm{\alpha}}\xi^b
-
\xi^bB_J^{\bm{\alpha}}{}_{\dot ab}
\partial_{\bm{\alpha}}\bar\xi^{\dot a}
}
{(\bar\xi^{\dot c}\sigma_{0\dot cd}\xi^d)^{1/2}}
\end{equation}
where $B_J^{\bm{\alpha}}{}_{\dot ab}$ are some constants and
summation is carried out over the spinor indices
$\dot a=\dot 1,\dot 2$, $b=1,2$,
and over $\bm{\alpha}=0,1,2,3$. Here
we use bold type to indicate relativistic
notation, when time $x^0$ is viewed as one of the coordinates in
(1+3)-dimensional spacetime.

Note that for given $J$ and $\bm{\alpha}$
the $2\times2$ matrices $B_J^{\bm{\alpha}}{}_{\dot ab}$ are Hermitian.
This is because
each of these matrices is a linear combination with real coefficients
of the Pauli matrices $\sigma^\beta{}_{\dot ab}$,
$\beta=1,2,3$.

Our action is $S(\xi)=\int L(\xi)$ where for the sake of brevity we
dropped $dx^0dx^1dx^2dx^3$.
Substituting
(\ref{Separation of variables equation 3})
into
(\ref{Separation of variables equation 2})
and varying the spinor field $\xi$ we get
\[
\delta S(\xi)=
2\int\sum_{J=1}^{16}A_J
\left(
i\frac{
(\delta\bar\xi^{\dot a})B_J^{\bm{\alpha}}{}_{\dot ab}
\partial_{\bm{\alpha}}\xi^b
-
\xi^bB_J^{\bm{\alpha}}{}_{\dot ab}
\partial_{\bm{\alpha}}\delta\bar\xi^{\dot a}
}
{(\bar\xi^{\dot c}\sigma_{0\dot cd}\xi^d)^{1/2}}
-\frac{
(\delta\bar\xi^{\dot a})\sigma_{0\dot ab}\xi^b
}
{2\bar\xi^{\dot c}\sigma_{0\dot cd}\xi^d}
W_J
\right)
W_J
+\mathrm{c.c.}
\]
where we wrote down explicitly the terms with $\delta\bar\xi$ and
incorporated the terms with $\delta\xi$ into the ``$\mathrm{c.c.}$''
(complex conjugate term).
Integrating the term with
$\partial_{\bm{\alpha}}\delta\bar\xi^{\dot a}$ by parts
and taking out the common factor $\delta\bar\xi^{\dot a}$ we rewrite
the above formula as
\[
\delta S(\xi)=
2\int(\delta\bar\xi^{\dot a})
\sum_{J=1}^{16}A_J
\left[
\frac{
iW_JB_J^{\bm{\alpha}}{}_{\dot ab}
\partial_{\bm{\alpha}}\xi^b
}
{(\bar\xi^{\dot c}\sigma_{0\dot cd}\xi^d)^{1/2}}
-\frac{
W_J^2\sigma_{0\dot ab}\xi^b
}
{2\bar\xi^{\dot c}\sigma_{0\dot cd}\xi^d}
+i\partial_{\bm{\alpha}}
\left(
\frac{
W_JB_J^{\bm{\alpha}}{}_{\dot ab}\xi^b
}
{(\bar\xi^{\dot c}\sigma_{0\dot cd}\xi^d)^{1/2}}
\right)
\right]
+\mathrm{c.c.}
\]
Comparing this formula with formula
(\ref{Euler--Lagrange equation equation 1})
we conclude that the spinor field $F$ appearing in the latter
is given by formula
\begin{equation}
\label{Separation of variables equation 4}
F_{\dot a}=
2\sum_{J=1}^{16}A_J
\left[
\frac{
iW_JB_J^{\bm{\alpha}}{}_{\dot ab}
\partial_{\bm{\alpha}}\xi^b
}
{(\bar\xi^{\dot c}\sigma_{0\dot cd}\xi^d)^{1/2}}
-\frac{
W_J^2\sigma_{0\dot ab}\xi^b
}
{2\bar\xi^{\dot c}\sigma_{0\dot cd}\xi^d}
+i\partial_{\bm{\alpha}}
\left(
\frac{
W_JB_J^{\bm{\alpha}}{}_{\dot ab}\xi^b
}
{(\bar\xi^{\dot c}\sigma_{0\dot cd}\xi^d)^{1/2}}
\right)
\right]
\end{equation}
where the $W_J$ are, in turn, given by formula
(\ref{Separation of variables equation 3}).

If we now substitute the plane wave
(\ref{Plane wave solutions equation 1})
into formula (\ref{Separation of variables equation 3})
we get
\begin{equation}
\label{Separation of variables equation 5}
W_J=
\frac{2
\bar\zeta^{\dot a}B_J^{\bm{\alpha}}{}_{\dot ab}
\mathbf{p}_{\bm{\alpha}}\zeta^b}
{(\bar\zeta^{\dot c}\sigma_{0\dot cd}\zeta^d)^{1/2}}\,.
\end{equation}
Note that the above $W_J$ are constant
(do not depend on $\mathbf{x}$), which simplifies the next step:
substituting
(\ref{Plane wave solutions equation 1})
into
(\ref{Separation of variables equation 4})
and dividing through by the common factor
$e^{-i\mathbf{p}\cdot\mathbf{x}}$
we get
\begin{equation}
\label{Separation of variables equation 6}
e^{i\mathbf{p}\cdot\mathbf{x}}F_{\dot a}=
2\sum_{J=1}^{16}A_J
\left[
\frac{
2W_JB_J^{\bm{\alpha}}{}_{\dot ab}
\mathbf{p}_{\bm{\alpha}}\zeta^b
}
{(\bar\zeta^{\dot c}\sigma_{0\dot cd}\zeta^d)^{1/2}}
-\frac{
W_J^2\sigma_{0\dot ab}\zeta^b
}
{2\bar\zeta^{\dot c}\sigma_{0\dot cd}\zeta^d}
\right].
\end{equation}

The remarkable feature of formula
(\ref{Separation of variables equation 6})
is that its RHS is constant,
i.e.~it does not depend on~$\mathbf{x}$.
This completes the proof of Lemma~\ref{lemma 1}.

Let us now substitute the plane wave
(\ref{Plane wave solutions equation 1})
directly into our Lagrangian density
(\ref{Separation of variables equation 2}).
Our Lagrangian density takes the form
\begin{equation}
\label{Separation of variables equation 7}
L(\zeta;\mathbf{p})=\sum_{J=1}^{16}A_JW_J^2
\end{equation}
where the $W_J$ are given by formula
(\ref{Separation of variables equation 5}).
The Lagrangian density (\ref{Separation of variables equation 7})
does not depend on $\mathbf{x}$.
The dynamical variable in this Lagrangian density is the constant
2-component complex spinor $\zeta$, whereas the relativistic 4-momentum
$\mathbf{p}$ plays the role of a parameter.
Varying the spinor $\zeta$ we get
\[
\delta L(\zeta;\mathbf{p})=
2\sum_{J=1}^{16}A_J
\left(
\frac{
2(\delta\bar\zeta^{\dot a})B_J^{\bm{\alpha}}{}_{\dot ab}
\mathbf{p}_{\bm{\alpha}}\zeta^b
}
{(\bar\zeta^{\dot c}\sigma_{0\dot cd}\zeta^d)^{1/2}}
-\frac{
(\delta\bar\zeta^{\dot a})\sigma_{0\dot ab}\zeta^b
}
{2\bar\zeta^{\dot c}\sigma_{0\dot cd}\zeta^d}
W_J
\right)
W_J
+\mathrm{c.c.}
\]
Comparing this formula with formula
(\ref{Plane wave solutions equation 16})
we conclude that the constant spinor $G$ appearing in the latter
is given by formula
\begin{equation}
\label{Separation of variables equation 8}
G_{\dot a}=
2\sum_{J=1}^{16}A_J
\left[
\frac{
2W_JB_J^{\bm{\alpha}}{}_{\dot ab}
\mathbf{p}_{\bm{\alpha}}\zeta^b
}
{(\bar\zeta^{\dot c}\sigma_{0\dot cd}\zeta^d)^{1/2}}
-\frac{
W_J^2\sigma_{0\dot ab}\zeta^b
}
{2\bar\zeta^{\dot c}\sigma_{0\dot cd}\zeta^d}
\right].
\end{equation}

It remains to observe that the right-hand sides of
formulae
(\ref{Separation of variables equation 6})
and
(\ref{Separation of variables equation 8})
are the same.
This completes the proof of Lemma~\ref{lemma 2}.

\end{document}